\documentclass[pre, twocolumn, showpacs, preprintnumbers]{revtex4}

\usepackage{amssymb}
\usepackage{amsmath}
\usepackage{graphicx}
\usepackage{dcolumn}
\usepackage{bm}
\usepackage{txfonts}
\DeclareMathAlphabet{\mathpzc}{OT1}{pzc}{m}{it}
\DeclareMathAlphabet{\mathcalligra}{T1}{calligra}{m}{n}
\begin{document}

\preprint{APS/123-QED}

\title{Preliminary test of time-convolutionless mode-coupling theory\\ based on the Percus-Yevick static structure factor for hard spheres}
\author{Yuto Kimura$^1$ and Michio Tokuyama$^2$}
\address{$^1$Department of Mechanical Engineering, Hachinohe National College of Technology, Hachinohe 039-1192, Japan\\
$^2$Institute of Multidisciplinary Research for Advanced Materials, Tohoku University, Sendai 980-8577, Japan}

\date{\today}

\begin{abstract}
In order to investigate how the time-convolutionless mode-coupling theory (TMCT) recently proposed by Tokuyama can improve the critical point predicted by the ideal mode-coupling theory (MCT), the TMCT equations are numerically solved based on the Percus-Yevick static structure factor for hard spheres as a preliminary test. Then, the full numerical solutions are compared with those of MCT for different physical quantities, such as intermediate scattering functions and diffusion coefficients. Thus, the ergodic to nonergodic transition predicted by MCT is also found at the critical volume fraction $\phi_c$ which is higher than that of MCT. Here $\phi_c$ is given by $\phi_c\simeq 0.5817$ at $q_c\sigma_d=40$ and 0.5856 at $q_c\sigma_d=20$ for TMCT, while $\phi_c\simeq 0.5159$ at $q_c\sigma_d=40$ and 0.5214 at $q_c\sigma_d=20$ for MCT, where $q_c$ is a cutoff of wave vector and $\sigma_d$ a particle diameter. The same two-step relaxation process as that predicted by MCT is also discussed. 
\end{abstract}

\maketitle

In order to discuss the dynamics of supercooled liquids, the so-called ideal mode-coupling theory (MCT) has been proposed by Bengtzelius, G\"{o}tze, and Sj\"{o}lander \cite{ben84}, and independently by Leutheusser \cite{leu84}. The MCT equations for the intermediate scattering function $F_{\alpha}(q,t)$ have been numerically solved for various glass-forming systems \cite{fuch,fuch2,fuchs,fran,phd,win,chong,voigt03,got03,sza,foffi,voigt04,flenn,voigt06,toku08,got09,voigt10,narumi11,voigt11}, where $\alpha=c$ stands for collective case and $\alpha=s$ for self case. Although the MCT full numerical solutions show an ergodic to non-ergodic transition at a critical temperature $T_c$ ( or a critical volume fraction $\phi_c$), $T_c$ (or $\phi_c)$ is always much higher (or lower) than the thermodynamic glass transition temperature $T_g$ (or $\phi_g$), which is commonly defined by a crossover point seen in an enthalpy-temperature line \cite{debe}. In order to overcome this high $T_c$ problem, Tokuyama \cite{toku14} has recently proposed the time-convolutionless MCT (TMCT) equations for $F_{\alpha}(q,t)$ by employing exactly the same formulation as that used in MCT, except that the time-convolutionless type projection operator method \cite{toku75} is applied for the density instead of the convolution type \cite{mori65} for the density and the current. Then, in the previous paper \cite{toku141} it has been shown within a simplified model proposed by MCT that there also exist non-zero long-time solutions for $T\leq T_c$ where $T_c$ is much lower than that of MCT. In the present paper, therefore, as a preliminary test of TMCT, we solve the TMCT equations numerically based on the Percus-Yevick (PY) static structure factor for hard spheres \cite{PY} under exactly the same conditions as those employed in the previous calculations of the MCT equations \cite{chong}. Thus, we show that $\phi_c$ is much higher than that of MCT.

We consider the three-dimensional equilibrium glass-forming system, which consists of $N$ particles with mass $m$ and diameter $\sigma_d$ in the total volume $V$ at temperature $T$. We define the intermediate scattering function by $F_{\alpha}(q,t)=\langle \rho_{\alpha}(\bm{q},t)\rho_{\alpha}(\bm{q},0)^*\rangle$ with the collective density fluctuation $\rho_c(\bm{q},t)=N^{-1/2}[\sum_{j=1}^N\rho_s(\bm{q},t)-N\delta_{\bm{q},0}]$ and the self density fluctuation $\rho_s(\bm{q},t)=e^{i\bm{q}\cdot\bm{X}_j(t)}$, where $\bm{X}_j(t)$ denotes the position vector of the $j$th particle at time $t$ and $q=|\bm{q}|$. Since the density fluctuations $\rho_{\alpha}(\bm{q},t)$ are macroscopic physical quantities, we set $q\leq q_c$, where the inverse cutoff $q_c^{-1}$ is longer than a linear range of the intermolecular force but shorter than a semi-macroscopic length and is in general fixed so that the numerical solutions coincide with the simulation results at least in a liquid state. Here $F_c(q,0)=S_c(q)=S(q)$ and $F_s(q,0)=S_s(q)=1$, where $S(q)$ is a static structure factor. As shown in the previous papers Refs. \cite{toku14,toku141}, the TMCT equations are then given by 
\begin{equation}
F_{\alpha}(q,t)=S_{\alpha}(q)\exp[-K_{\alpha}(q,t)],\label{F}
\end{equation}
\begin{eqnarray}
\frac{\partial^2 K_{\alpha}(q,t)}{\partial t^2}&=&\frac{q^2v_{th}^2}{S_{\alpha}(q)}-\gamma_{\alpha}\frac{\partial K_{\alpha}(q,t)}{\partial t}\nonumber\\
&-&\int_0^t\Delta\varphi_{\alpha}(\bm{q},t-\tau)\frac{\partial K_{\alpha}(q,\tau)}{\partial \tau}d\tau\label{psi}
\end{eqnarray}
with the nonlinear memory function $\Delta\varphi_{\alpha}(\bm{q},t)$ given by
\begin{equation}
\Delta\varphi_{\alpha}(\bm{q},t)=\frac{v_{th}^2}{2^{n_{\alpha}}\rho}\int_<\frac{d\bm{k}}{(2\pi)^3}v_{\alpha}(\bm{q},\bm{k})^2F_c(k,t)F_{\alpha}(|\bm{q}-\bm{k}|,t),\label{memory}
\end{equation}
where $\gamma_{\alpha}$ is a positive constant and $\int_<$ denotes the sum over wave vectors $\bm{k}$ whose magnitudes are smaller than a cutoff $q_c$. Here the initial conditions for $K_{\alpha}$ are given by $K_{\alpha}(q,t=0)=dK_{\alpha}(q,t)/dt|_{t=0}=0$. The vertex amplitude $v_{\alpha}(\bm{q},\bm{k})$ is given by $
v_{\alpha}(\bm{q},\bm{k})=\hat{\bm{q}}\cdot\bm{k}c(k)+n_{\alpha}\hat{\bm{q}}\cdot(\bm{q}-\bm{k})c(|\bm{q}-\bm{k}|)$,
where $c(k)=1-1/S(k)$, $n_c=1$, $n_s=0$, $\rho=N/V$, $v_{th}=(k_BT/m)^{1/2}$, and $\hat{\bm{q}}=\bm{q}/q$. On the other hand, the MCT equation is given by \cite{ben84}
\begin{eqnarray}
\frac{\partial^2F_{\alpha}(q,t)}{\partial t^2}&=&-\frac{q^2v_{th}^2}{S_{\alpha}(q)}F_{\alpha}(q,t)-\gamma_{\alpha}\frac{\partial F_{\alpha}(q,t)}{\partial t}\nonumber\\
&-&\int_0^t\Delta\varphi_{\alpha}(\bm{q},t-\tau)\frac{\partial F_{\alpha}(\bm{q},\tau)}{\partial \tau}d\tau.\label{mct}
\end{eqnarray}
Here we note that Eq. (\ref{psi}) has a form similar to Eq. (\ref{mct}).

\begin{figure}
\begin{center}
\includegraphics[width=8.5cm]{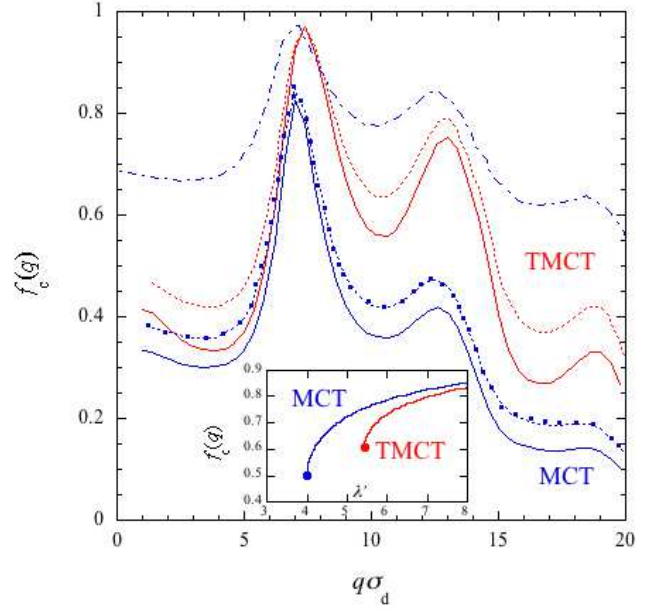}
\end{center}
\caption{(Color online) A plot of the Debye-Waller factor $f_c(q)$ versus $q$. The solid lines indicate the numerical results for $f_c^c(q)$ at $q_c\sigma_d=20$ and the dotted lines at $q_c\sigma_d=40$ and the symbols $(\bullet)$ the numerical results at $q_c\sigma_d=40$ from Ref. \cite{voigt04}. The dot-dashed line indicates the nonergodicity parameter $f_c(q)$ of MCT for $\phi=0.55$ at $q_c\sigma_d=40$ from Ref. \cite{ben84}. The insert shows $f_c(q_m)$ versus $\lambda'$ for a simplified model and the symbols indicate the values at $\lambda_c'$. }
\label{dw}
\end{figure}

The most important prediction of MCT is the ergodic to non-ergodic transition at a critical temperature $T_c$, below which the solution $F_{\alpha}(q,t)$ reduces to a non-zero value $f_{\alpha}(q)$ for long times, which is the so-called nonergodicity parameter. In fact, from Eq. (\ref{mct}), one can find \cite{ben84}
\begin{equation}
f_{\alpha}(q)=\lim_{t\rightarrow \infty}\frac{F_{\alpha}(q,t)}{S_{\alpha}(q)}=\frac{\mathpzc{F}_{\alpha}(q)}{1+\mathpzc{F}_{\alpha}(q)}\label{mctfq}
\end{equation}
with the long-time limit of the memory function
\begin{equation}
\mathpzc{F}_{\alpha}(q,f_c,f_{\alpha})
=\frac{1}{2^{n_{\alpha}}(2\pi)^3}\int_< d\bm{k}V_{\alpha}^{(2)}(q, k, |\bm{q}-\bm{k}|)f_c(k)f_{\alpha}(|\bm{q}-\bm{k}|),\label{Ga1}
\end{equation}
where the vertex $V_{\alpha}^{(2)}$ is given by
\begin{equation}
V_{\alpha}^{(2)}(q, k, |\bm{q}-\bm{k}|)=S_{\alpha}(q)S_c(k)S_{\alpha}(|\bm{q}-\bm{k}|)v_{\alpha}(\bm{q},\bm{k})^2/(\rho q^2).\label{V}
\end{equation}
As shown in the previous papers \cite{toku14,toku141}, this prediction also holds for TMCT.  In fact, from Eqs. (\ref{F}) and (\ref{psi}) the non-zero solution is given by
\begin{equation}
f_{\alpha}(q)=\exp\left[- \frac{1}{\mathpzc{F}_{\alpha}(q)}\right].\label{fq}
\end{equation}
In order to estimate how the critical point obtained by Eq. (\ref{fq}) is different from that by Eq. (\ref{mctfq}), it is convenient to employ the simplified model discussed by Bengtzelius et al \cite{ben84}. Then, one can write $S(q)$ as $S(q)=1+A\delta(q-q_m)$, where $A$ is a positive constant to be determined and $q_m$ a wave vector of the first peak of $S(q)$. Then, one can write Eq. (\ref{Ga1}) as $\mathpzc{F}_c(q_m)=\lambda' f_c(q_m)^2$, where the coupling parameter $\lambda'$ is given by $\lambda'=q_mA^2S(q_m)/(8\pi^2\rho)$. Use of Eqs. (\ref{mctfq}) and (\ref{Ga1}) then leads to the critical coupling parameter $\lambda_c'=4$ and $f_c^c=1/2$, while use of Eqs. (\ref{Ga1}) and (\ref{fq}) leads to $\lambda_c'=2e(\simeq 5.43656)$, $K_c^c=1/2$, and $f_c^c=e^{-K_c^c}=e^{-1/2}(\simeq 0.60653)$. Thus, the critical coupling parameter of TMCT is larger than that of MCT (see the insert in Fig. \ref{dw}). Hence this suggests that the critical temperature $T_c$ (or the critical volume fraction $\phi_c$) of TMCT is much lower (or higher) than that of MCT. Therefore, we next check this by solving the TMCT equations numerically based on the PY static structure factor.

\begin{figure}
\begin{center}
\includegraphics[width=8.5cm]{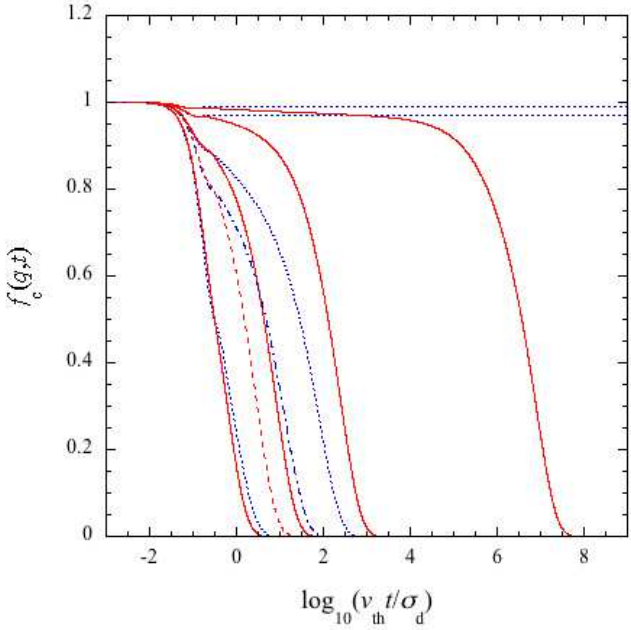}
\end{center}
\caption{(Color online) A plot of $f_c(q,t)$ versus scaled time $t/t_0$ for different volume fractions at $q_c\sigma_d=40$, where $q\sigma_d=7.4$. The solid lines indicate the TMCT results and the dotted lines the MCT results for $\phi=$0.40, 0.50, 0.55, and 0.58 from left to right. The dashed line indicates the TMCT results for $q_c\sigma_d=20$ at $\phi=0.5$ and the dot-dashed line for the MCT ones.}
\label{fcall}
\end{figure}

We now solve the TMCT equations numerically by using the PY static structure factor under the same conditions as those employed by Chong et al \cite{chong} to solve the MCT equations at $q_c\sigma_d=40$ and $\gamma_{\alpha}=0$. Here the MCT equations are also solved and the solutions are compared with the previous results obtained from Ref. \cite{chong,voigt04} to check whether the present calculations are correct or not. The control parameter is the volume fraction given by $\phi=\pi \sigma_d^3N/(6V)$. We put $\gamma_{\alpha}=0$ but take two different cutoffs as $q_c\sigma_d=20$ and 40 here. Then, we first find the critical volume fraction $\phi_c$ and the so-called Debye-Waller factor $f_c^c(q)$ for TMCT by solving Eqs. (\ref{Ga1}) and (\ref{fq}) and also for MCT by solving Eqs. (\ref{mctfq}) and (\ref{Ga1}). In Fig. \ref{dw}, the critical Debye-Waller factor $f_c^c(q)$ is plotted versus $q\sigma_d$ for different cutoffs $q_c$, where the values of $\phi_c$ are listed in Table \ref{table-1}.
\begin{table}
\caption{$\phi_c$ for MCT and TMCT at different wave vector cutoff $q_c$.}
\begin{center}
\begin{tabular}{ccc}
\hline
Theory & \multicolumn{2}{c}{$\phi_c$}\\
 & $q_c\sigma_d=$20& 40\\
\hline
MCT &0.5214& 0.5159\\
TMCT &0.5856&0.5817\\
\hline
\end{tabular}
\end{center}
\label{table-1}
\end{table}
Thus, it is shown that since $\phi_c$ of TMCT is much higher than that of MCT, $f_c^c$ of TMCT is larger than that of MCT. It is also shown that in both theories $f_c^c(q)$ at $q_c\sigma_d=40$ is larger than that at $q_c\sigma_d=20$. However, we should mention here that at higher volume fractions, the nonergodicity parameter of MCT is always larger than that of TMCT. In fact, for comparison the nonergodicity parameter of MCT for $\phi=0.55$ is also plotted at $q_c\sigma_d=40$ (see also the insert in Fig. \ref{dw}). In order to check the present MCT numerical solutions, the MCT numerical solutions obtained for the PY static structure factor at $q_c\sigma_d=40$ by Voigtmann et al \cite{voigt04} are also shown. The present results agree with them within error. In Fig. \ref{fcall}, the scaled collective-intermediate scattering function $f_c(q,t)(=F_c(q,t)/S(q))$ is plotted versus scaled time $v_{th}t/\sigma_d$ at $q_c\sigma_d=40$ for different volume fractions, where $q\sigma_d=7.4$. For comparison, the numerical results for $q_c\sigma_d=20$ are also plotted at $\phi=0.5$ by the dashed line for TMCT and the dot-dashed line for MCT.

\begin{figure}[t]
\begin{center}
\includegraphics[width=8.5cm]{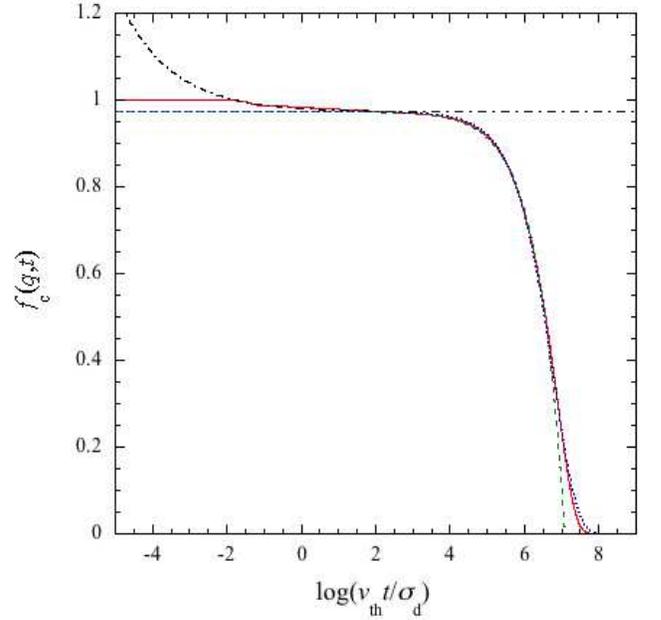}
\end{center}
\caption{(Color online) A plot of $f_c(q,t)$ versus scaled time $t/t_0$ for $\phi=0.58$ (or $1-\phi/\phi_c=2.92\times 10^{-3}$) at $q_c\sigma_d=40$, where $q\sigma_d=7.4$. The solid line indicates the TMCT results and the dot-dashed line the critical decay, the dashed line the von Schweidler decay, and the dotted line the KWW decay, where $\lambda=0.735$ ($a=0.312$, $b=0.583$), $\beta=0.70$, and $f_c^c=0.973$.}
\label{power}
\end{figure}
We next discuss the asymptotic behavior of $f_c(q,t)$ in each time stage. As demonstrated in Refs. \cite{got84,got91}, MCT shows that $f_c(q,t)$ obeys a characteristic two-step relaxation process at the so-called $\beta$-relaxation stage [$\beta$] near the critical point. By introducing the Laplace transform $f_c[q,z]$ of $f_c(q,t)$ by $f_c[q,z]=\mathcal{L}[f_c(q,t)][z]:= \int_0^{\infty}e^{-zt}f_c(q,t)dt$, the long-time dynamics is then determined from Eq. (\ref{mct}) as
\begin{equation}
\frac{zf_c[q,z]}{1-zf_c[q,z]}=z\mathcal{L}[\mathpzc{F}_c(q,f_c(t),f_c(t))][z].\label{mctz}
\end{equation}
Following MCT \cite{got91}, one can split $f_c(q,t)$ into the trivial asymptotic part and the a non-trivial part $G$; 
\begin{equation}
f_c(q,t)=f_c^c(q)+h_qG(t),\;\;\; zf_c[q,z]=f_c^c(q)+zh_qG[z]\label{split}
\end{equation}
with $h_q=(1-f_c^c(q))^2e_q^c$, where $e_q^c$ is an appropriately normalized right eigenvector of the stability matrix $C_{qk}=(\partial \mathpzc{F}_c/ \partial f_c(k))(1-f_c^c(k))^2$ at $\phi_c$. From Eqs. (\ref{mctz}) and (\ref{split}), one can then find near $\phi_c$
\begin{equation}
\sigma+\lambda\{z\mathcal{L}[G(t)^2][z]\}-\{zG[z]\}^2=0, \label{sol}
\end{equation}
where $\sigma$ is a separation parameter and $\sigma=0$ at $\phi=\phi_c$. Here $\lambda$ is the so-called exponent parameter given by $\lambda=(1/2)\sum_{q,k,p}\hat{e}_q^cV_c^{(2)}(q,k,p)h_kh_p$, where $\hat{e}_q^c$ is a left eigenvector defined by $\sum_q\hat{e}_q^ce_q^c=1$. Thus, use of Eq. (\ref{sol}) leads to two different power-law decays for $G(t)$ near $\phi_c$; the so-called critical decay at a fast $\beta$ stage
\begin{equation}
G(t)=|\sigma|^{1/2}(t_{\sigma}/t)^a,\;\;\; t_0\ll t\leq t_{\sigma}, \label{critical}
\end{equation}
and the so-called von Schweidler decay at a slow $\beta$ stage
\begin{equation}
G(t)=-(t/t_{\sigma}')^b, \;\;\; t_{\sigma}\leq t \ll t_{\sigma}', \label{von}
\end{equation}
where $t_0$, $t_{\sigma}$, and $t_{\sigma}'$ are characteristic times. Here the time exponents $a$ and $b$ are determined by $\Gamma[1-a]^2/\Gamma[1-2a]=\Gamma[1+b]^2/\Gamma[1+2b]=\lambda$. For the details of parameters the reader is referred to Ref. \cite{got91}. For the PY model, $\lambda$ is calculated as $\lambda=0.735$ at $q_c\sigma_d=40$, leading to $a=0.312$ and $b=0.583$ \cite{fran}. On the other hand, in TMCT use of Eqs. (\ref{F}) and (\ref{psi}) leads to
\begin{equation}
\frac{1}{zK_c[q,z]}=z\mathcal{L}[\mathpzc{F}_c(q,f_c(t),f_c(t))][z].\label{tmctz}
\end{equation}
From Eqs. (\ref{F}) and (\ref{split}), one can find, up to lowest order in $h_q$,
\begin{equation}
K_c(q,t)=K_c^c(q)-h_qG(t)/f_c^c,\;\; zK_c[q,z]=K_c^c(q)-zh_qG[z]/f_c^c,\label{split2}
\end{equation}
where $K_c^c=-\ln(f_c^c)$. One can then directly apply the same formulation as that employed by MCT to Eq. (\ref{tmctz}) near $\phi_c$. In fact, from Eqs. (\ref{tmctz}) and (\ref{split2}) one can obtain Eq. (\ref{sol}) under the condition $f_c^c=e^{-K_c^c}\simeq 1-K_c^c$. Hence $G(t)$ also obeys Eqs. (\ref{critical}) and (\ref{von}). Use of Eqs. (\ref{F}) and (\ref{split2}) thus leads to the same two-step relaxations for $f_c(q,t)$ as those of MCT, up to lowest order. Since $\lambda$ is determined at $\phi_c$, $\lambda$ of TMCT must have the same value as that of MCT. This can be easily checked within a simplified model. Since $\lambda$ of MCT is known for the PY model, one can also use it for TMCT to check this. In fact, in Fig. \ref{power} the TMCT results are shown to be well described by the same value of $\lambda$ as that of MCT near $\phi_c$. Finally, at the so-called $\alpha$-relaxation stage after the $\beta$ stage, $f_c(q,t)$ is also shown to obey the Kohlrausch-Williams-Watts (KWW) function, i.e., $f_c(q,t)=f_c^c(q)\exp[-(t/\tau_{\alpha})^{\beta}]$ with a stretched exponent $\beta$ and an $\alpha$-relaxation time $\tau_{\alpha}$.  Thus, the numerical solutions of TMCT are shown to be well described by the same asymptotic laws as those obtained by MCT.

\begin{figure}
\begin{center}
\includegraphics[width=8.5cm]{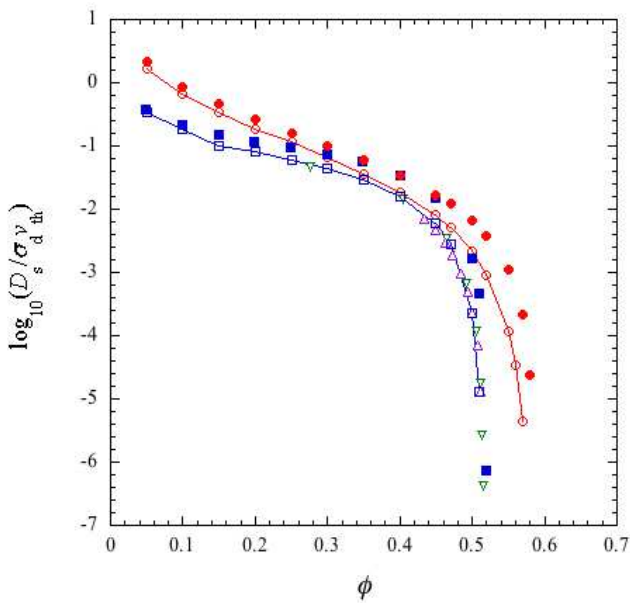}
\end{center}
\caption{(Color online) A log plot of $D_s$ versus $\phi$. The filled symbols indicate the numerical results obtained by TMCT and MCT at $q_c\sigma_d=20$ and the open symbols at $q_c\sigma_d=40$. The symbols $(\circ)$ indicate the TMCT results and $(\Box)$ the MCT results. The symbols $(\triangledown)$ indicate the MCT results obtained by Chong et al from Ref. \cite{chong} and $(\triangle)$ those obtained by Voigtmann et al from Ref. \cite{voigt04}. The solid lines are guides to eyes for $q_c\sigma_d=40$.}
\label{Ds}
\end{figure}
In order to compare the dynamics of a tagged particle in TMCT with that in MCT, we finally discuss the long-time self-diffusion coefficient $D_s$, which is given in both theories by
\begin{equation}
D_s=\frac{1}{q^2}\frac{dK_s(q=0,t)}{dt}|_{t=\infty}=\frac{v_{th}^2}{\gamma_s+\int_0^{\infty}\Delta\varphi_s(\bm{q}=0,\tau)d\tau}.\label{ds}
\end{equation}
In Fig. \ref{Ds}, $D_s$ is plotted versus $\phi$ at $\gamma_s=0$ for different cutoffs. Thus, $D_s$ of TMCT is shown to be always larger than that of MCT. In both theories, $D_s$ for $q_c\sigma_d=20$ is also shown to be larger than that for $q_c\sigma_d=40$. This is simply because the magnitude of the nonlinear memory function $\Delta\varphi_s(\bm{q}=0,t)$ in Eq. (\ref{ds}) decreases as $q_c$ decreases. This is consistent with the fact that $\phi_c$ for $q_c\sigma_d=20$ is always higher than that for $q_c\sigma_d=40$. The numerical calculations of the MCT equations based on the PY static structure factor have been already done at $q_c\sigma_d=40$ by the other authors \cite{chong,voigt04}. For comparison, therefore, their results are also shown in Fig. \ref{Ds}. All the MCT results for $q_c\sigma_d=40$ coincide with each other within error. Here we note that in both theories $D_s/\sigma_d v_{th}$ becomes larger than one for lower volume fractions because of $\gamma_s=0$. This is also seen in the MCT results obtained by Fuchs \cite{fuchs}, where the Verlet-Weis approximation for $S(q)$ has been used.

In this paper, we have solved not only the TMCT equations but also the MCT equations numerically by using the PY static structure factor under the same conditions as employed in the previous works and compared the TMCT results with the MCT results. We have first checked whether $\phi_c$ of MCT at $q_c\sigma_d=40$ coincides with the common value 0.516 obtained in the previous MCT calculations or not (see Table \ref{table-1}). Then, we have shown that in both theories all the numerical results depend on the cutoff $q_c$. In fact, for smaller $q_c$ $\phi_c$ is higher and $f_c$ is smaller in both theories. Thus, we have shown that $\phi_c$ of TMCT is much higher than that of MCT, irrespectively of the magnitude of $q_c$. We have also shown that there exists the same two-step relaxation process in a $\beta$ stage as that discussed in MCT near $\phi_c$. In order to check whether TMCT can describe the dynamics of supercooled liquids reasonably well or not, the TMCT equations must be solved numerically by using the static structure factor obtained from the simulations and the experiments. This will be discussed elsewhere.

The authors wish to thank E. Flenner, T. Narumi, G. Szamel, and Th. Voigtmann for their kind advices in calculating the MCT equations. One of the authors (Y.K.) also thanks F. Takeo, A. Miyamoto, and N. Hatakeyama for giving him a chance of sabbatical and their encouragement. This work was partially supported by High Efficiency Rare Elements Extraction Technology Area, IMRAM, Tohoku University, Japan.

\end{document}